\documentclass[a4paper,11pt]{article}

\input{xy}
\xyoption{all}
\xyoption{poly}
\usepackage[all]{xy}
\usepackage{amsmath,amsthm,amsfonts,amssymb,stmaryrd,fancyhdr,mathtools}
\usepackage{latexsym}
\usepackage{color}
\usepackage{graphicx}
\usepackage{float}
\usepackage{fullpage}
\usepackage{mdframed}
\usepackage[font=small,labelfont=small]{caption}
\usepackage[T1]{fontenc}
\usepackage[utf8]{inputenc}
\usepackage{authblk}

\def\beq{\begin{equation}}
\def\eeq{\end{equation}}
\def\beqa{\begin{eqnarray}}
\def\eeqa{\end{eqnarray}}

\title{More $AdS_3$ correlators}
\author{Yago~Cagnacci\thanks{yago@iafe.uba.ar} \,}
\author{\, Sergio~M.~Iguri\thanks{siguri@iafe.uba.ar}}
\affil{Instituto de Astronom\'{\i}a y F\'{\i}sica del Espacio (CONICET-UBA)

C.~C.~67, Suc.~28, 1428 Buenos Aires, Argentina}

\begin{document}

\maketitle

\begin{abstract}

We compute three-point functions for the $SL(2,\mathbb R)$-WZNW model. After reviewing the case of the two-point correlator, we compute spectral flow preserving and nonpreserving correlation functions in the space-time picture involving three vertex operators carrying an arbitrary amount of spectral flow. When only one or two insertions have nontrivial spectral flow numbers, the method we employ allows us to find expressions without any constraint on the spin values. Unlike these cases, the same procedure restrains the possible spin configurations when three vertices belong to nonzero spectral flow sectors. We perform several consistency checks on our results. In particular, we verify that they are in complete agreement with previously computed correlators involving states carrying a single unit of spectral flow.

\end{abstract}


\bigskip

\section{Introduction}

The $SL(2,\mathbb R)$-WZNW model has many physical applications, ranging from gravity and string theory \cite{Seiberg:1990eb,Witten:1988hc,Witten:1991yr,Callan:1991at,Mukhi:1993zb,Ghoshal:1995wm,Ooguri:1995wj,Giveon:1999px,Giveon:1999tq,Banados:1992wn} to condensed-matter physics \cite{Zirnbauer:1996zz,Zirnbauer:1999ua,Bhaseen:1999nm,Kogan:1999hz}. The model has been intensively studied since the AdS/CFT correspondence was conjectured \cite{Maldacena:1997re,Witten:1998qj,Aharony:1999ti}, as it describes the world sheet of a string propagating in $AdS_3$ with a background NS-NS $2$-form $B$ field. So far, this is one of the few known schemes in which Maldacena's conjecture can be explored beyond the supergravity approximation with complete control over the world-sheet theory. The interest in models with a $SL(2,\mathbb R) \times SL(2,\mathbb R)$ global symmetry has been recently renewed within the context of integrability \cite{Beisert:2010jr} and $AdS_3$ gravity \cite{Sundborg:2013bya}.

In a WZNW model with a compact underlying symmetry group, the spectrum is built upon representations of the corresponding zero-modes algebra \cite{Gepner:1986wi}. Once all the vectors in these representations are assumed to be annihilated by those modes with positive frequency, representations of the full current algebra are constructed by acting on them with the negative-degree modes. Unitarity further restricts the possible configurations. When this standard procedure is implemented in the $SL(2,\mathbb R)$-WZNW model, which is a nonrational CFT, for those zero-modes representations with energy bounded below, the absence of negative norm states in the physical spectrum of the string can be reached if the allowed values of the spin have an upper bound \cite{Balog:1988jb,Dixon:1989cg,Bars:1990rb,Hwang:1998tr,Evans:1998qu,Petropoulos:1999nc}. This bound implies a coupling independent restriction on the masses of the physical states that originally raised doubts about whether a no-ghost theorem could be proven for strings in $AdS_3$. Moreover, the spectrum generated so far is empty of long string states, {\em i.e.}, finite energy states corresponding to strings stretched closed to the boundary of $AdS_3$ \cite{Maldacena:1998uz,Seiberg:1999xz}
.

Both problems were solved in \cite{Maldacena:2000hw}. In this reference, a spectrum involving unbounded energy representations was proposed for the $SL(2,\mathbb R)$-WZNW model and the no-ghost theorem based on this spectrum was proven. The case in which the target space is the universal cover of the $SL(2,\mathbb R)$ group manifold has always been considered. In \cite{Israel:2003ry} (see also \cite{Maldacena:2000kv}) the proposal was verified by computing the modular invariant partition function for a string in $AdS_3 \times \mathcal M$, where $\mathcal M$ is a compact space represented by a unitary CFT on the world sheet.

A key ingredient for generating the full spectrum as in \cite{Maldacena:2000hw} is the concept of spectral flow. The spectral flow automorphisms constitute a family of automorphisms of the current algebra labeled by an integer number $\omega$, the so-called spectral flow number, which, for some states, could be recognized as the amount of winding of the string in the angular direction of $AdS_3$. For WZNW models based on compact Lie groups, spectral flow relates standard representations, mapping a primary state of one into a current algebra descendant of another. Unlike the rational case, in the $SL(2,\mathbb R)$-WZNW model representations with different spectral flow numbers turn to be, generally, nonequivalent, the spectral flow automorphisms thus defining new representations from the conventional ones. These spectral flowed representations are generated by infinitely many affine quasi-primary fields and they have energies unbounded below, giving account of long string states.

Concerning the correlation functions of the model, those with only unflowed vertex operators are obtained from the correlators in its Euclidean counterpart, namely, the $H_3^+$-WZNW model \cite{Teschner:1997ft,Teschner:1999ug}, by means of a proper analytic continuation. More care must be taken, however, when dealing with amplitudes involving flowed insertions. There are two strategies for computing this kind of correlators, both exploiting the singular properties of the so-called spectral flow operator\footnote{We notice that an alternative efficient method to deal with Lorentzian $AdS_3$ is given in [34].}.

One of them was developed in \cite{Maldacena:2001km} and, according to it, the computation is completely performed in the original space-time basis, also called the $x$ basis for short. Roughly, the vertex operators associated with flowed states are expressed as unflowed vertices convoluted with spectral flow operators, the corresponding integrals being understood to hold inside any correlation function. This procedure was, then, used for determining the propagator of two states carrying a single unit of spectral flow each and the three-point function involving two unflowed states and one vertex with $\omega=1$. In \cite{Minces:2005nb} these results were generalized by computing three-point functions with only one unflowed state and two insertions with a unit of spectral flow and certain three-point correlators with all of their vertices in the $\omega=1$ sector. The main restraint of the method relies in the fact that the referred integral definition of a flowed vertex exists, so far, for operators with a single unit of spectral flow, the generalization for an arbitrary amount of spectral flow being still lacking.

The other strategy, 
the so-called FZZ method, was firstly presented in \cite{Fateev} based on the parafermionic vertices and the properties of their correlation functions. It does not suppose, {\em a priori}, any constraint on the value of $\omega$. Starting with a regular unflowed correlator, a spectral flow operator is inserted for each unit of spectral flow carried by each vertex. After Fourier-transforming the amplitude thus obtained to the $m$ basis, {\em i.e.}, the basis in which the Cartan generator of $SL(2,\mathbb R)$ is diagonal, the dependence on the ``unphysical'' insertion points is removed and the world-sheet dependence, adjusted. The computation eventually concludes when transforming back to the original space-time basis.

One of the consequences that can be read off from the FZZ procedure is that, up to their dependence on the world-sheet coordinates, correlation functions in the $m$ basis carrying the same total amount of spectral flow coincide. This fact can be exploited in order to reduce the number of inserted spectral flow operators to a minimum value. It follows, for instance, that not only amplitudes with unflowed vertices, but also spectral flow preserving correlators in the $m$ basis involving vertex operators in flowed frames, are determined by Fourier transforming the corresponding analytically continued expressions of the Euclidean model. In \cite{Becker:1993at,Giveon:1998ns,Kutasov:1999xu,Giribet:2000fy,Hosomichi:2001fm,Giveon:2001up,Giribet:2001ft,Satoh:2001bi,Minces:2007td,Giribet:2008yt,Iguri:2009cf,Baron:2008qf} it is possible to find spectral flow conserving two-, three- and four-point functions obtained following several techniques, in particular, using the free fields methods. The computation of correlation functions violating the conservation of the total spectral flow number is more involved since, as we have already said, they require the additional insertion of spectral flow operators. Spectral flow nonpreserving correlators are discussed in \cite{Maldacena:2001km,Giribet:2001ft,Iguri:2007af,Giribet:2011xf}.

Transforming the correlation functions in the $m$ basis back to the space-time picture is, by far, a nontrivial task. Nevertheless, when the affine symmetry of the current algebra fully dictates the functional dependence of the correlators on the space-time coordinates as, for example, for the two- and three-point functions, no integral transformation is needed and the FZZ recipe can be finally realized. In this paper we follow these ideas in order to compute three-point functions in the space-time representation with no restriction on the amount of spectral flow the states carry. Our results generalize, thus, those obtained in \cite{Maldacena:2001km,Minces:2005nb} in the $\omega=1$ sector.

The paper is organized as follows. In section \ref{sec2} we present some generalities of the $SL(2,\mathbb R)$-WZNW model and its Euclidean counterpart. We stress that by the former we mean the WZNW model whose target space is the universal cover of the $SL(2,\mathbb R)$ group manifold. After analyzing the spectra of both theories, the superselection rules for spectral flow violation are discussed and the expressions of the previously computed spectral flow conserving and nonconserving correlators with two and three insertion points are given. In section \ref{sec3}, after introducing the vertex operators associated with spectral flowed states, we review the computation of the propagator in the space-time picture as in \cite{Maldacena:2001km}. We also clarify some aspects concerning the notation, since ours differs from the one employed in this reference. In section \ref{sec4} we compute all nontrivial three-point functions with insertions carrying an arbitrary amount of spectral flow. When only two vertices belong to nontrivial spectral flow sectors, we find no constraint on the spin configurations. This is not the case when all the vertex operators involved have nonzero spectral flow number. We further discuss this matter. Finally, we present our conclusions.

\section{The $H_3^+$-WZNW and $SL(2,\mathbb R)$-WZNW models: some basics}
\label{sec2}

The elements of the hyperbolic space $H_3^+$ are the $2 \times 2$ Hermitian matrices with determinant equal to one. Since it can be realized as the right-coset space $SL(2,\mathbb C)/SU(2)$, the sigma model having $H_3^+$ as target space can accordingly be constructed as a coset of the $SL(2,\mathbb C)$-WZNW model by the right action of $SU(2)$. The Lagrangian is expressed in terms of the so-called Poincar\'e coordinates $(\phi,u,\bar u)$ as \cite{Gawedzki:1991yu}
\begin{equation}
\mathcal L = k \left( \partial \phi \bar{\partial} \phi + e^{2\phi} \partial \bar u \bar{\partial} u\right),
\end{equation}
where $k$ is the level of the model, related to the scalar curvature of $H_3^+$, $\partial=\partial/\partial z$ and $\bar{\partial}=\partial/\partial \bar{z}$, $(z,\bar z)$ being the (complex) coordinates of the world sheet. The corresponding action has a set of holomorphic and antiholomorphic conserved currents, $J^a(z)$ and $\bar{J}^a(\bar{z})$, with $a=3,\pm$, respectively, whose modes $J^a_n$ and $\bar{J}^a_n$ generate two commuting isomorphic $\mathfrak{sl}_2$-current algebras. The generators $J^a_n$ satisfy the following commutation relations,
\begin{equation}
\left[J^3_n,J^3_m\right]=-\frac{1}{2}kn\delta_{n+m,0}, \quad\quad \left[J^3_n,J^{\pm}_m\right]=\pm J^{\pm}_{n+m}, \quad\quad \left[J^-_n,J^+_m\right]=2 J^3_{n+m} + kn\delta_{n+m,0},
\end{equation}
and similarly the antiholomorphic modes $\bar{J}^a_n$.

Associated with the current algebra there are, as usual, two commuting Virasoro algebras constructed by means of the Sugawara procedure. Their generators are given by
\begin{equation}
L_m=\frac{1}{k-2} \sum_{n=1}^{\infty} \left(J^+_{m-n}J^-_{n} + J^-_{m-n}J^+_{n} -2 J^3_{m-n}J^3_{n}\right),
\end{equation}
for every $m \ne 0$, and
\begin{equation}
L_0=\frac{1}{k-2} \left[ \frac{1}{2}\left(J^+_0J^-_0+J^-_0J^+_0\right)-\left(J^3_0\right)^2 + \sum_{n=1}^{\infty} \left(J^+_{-n}J^-_{n} + J^-_{-n}J^+_{n} -2 J^3_{-n}J^3_{n}\right)\right],
\end{equation}
and they satisfy the commutation relation
\begin{equation}
[L_m,L_n] = (n-m) L_{n+m} + \frac{c}{12} n (n^2-1) \delta_{n+m,0},
\end{equation}
where the central charge is given by
\begin{equation}
c=\frac{3k}{k-2}.
\end{equation}
Analogous expressions hold for the antiholomorphic generators $\bar L_n$.

The space of states of the model is decomposed into a sum of irreducible unitary representations of the current algebra \cite{Teschner:1997ft,Teschner:1999ug}. They are parametrized by the spin $j = -1/2 + i\lambda$, with $\lambda \in \mathbb R_{>0}$, and are defined as follows. One considers a representation for the zero modes $J^a_0$ and $\bar{J}^a_0$, corresponding to a principal series of $SL(2,\mathbb C)$, and after requiring it to be annihilated by $J^a_n$ and $\bar{J}^a_n$ for every $n>0$, it is extended to a representation of the full current algebra by acting with its negative triangular part, namely, with the generators $J^a_n$ and $\bar{J}^a_n$ with $n<0$.

A concrete realization of the spectrum may be obtained by means of operators $\Phi_j(x|z)$ having the following characteristic OPEs with the currents,
\begin{equation}
\label{ope}
J^a(z)\Phi_j(x|w) \sim \frac{1}{z-w} D^a_j \Phi_j(x|w),
\end{equation}
where
\begin{equation}
D^-_j = -\partial_x, \quad\quad D^3_j = -x\partial_x+j, \quad\quad D^+_j = -x^2\partial_x+2jx,
\end{equation}
and the same for $\bar{J}^a(z)$ with similar expressions for $\bar{D}^a_j$. The labels $(x,\bar x)$ are coordinates that parametrize $S^2$, the boundary of $H_3^+$, which is the target space of the dual two-dimensional CFT \cite{Giveon:1998ns,Kutasov:1999xu}, so that we shall refer to this realization as the space-time picture. Notice that the operator $\Phi_j(x|z)$ is not only an affine primary but also a primary for the Sugawara-Virasoro algebra with conformal dimension
\begin{equation}
\Delta_0=-\frac{j(1+j)}{k-2}.
\end{equation}

In the semiclassical regime, the operator $\Phi_j(x|z)$ can be identified with the following conti\-nu\-um-normalizable function on $H_3^+$,
\begin{equation}
\Phi_j(x|z) = \frac{1+2j}{\pi} \left( \left( u - x \right)\left( \bar u - \bar x \right) e^{b \phi/\sqrt{2} } + e^{-b \phi/\sqrt{2} } \right)^{2j},
\end{equation}
where $b^2=(k-2)^{-1}$. Normal ordering does not not allow the quantum operator to get such a simple form. However it simplifies in the large-$\phi$ limit where the interaction vanishes, leading to
\begin{equation}
\Phi_j(x|z) = :e^{-\sqrt{2} b (1+j)\phi}:\delta^2(u-x)+B(j):e^{\sqrt{2} b j\phi}\left( u - x \right)^{2j}\left( \bar u - \bar x \right)^{2j}:.
\end{equation}
This expression fixes the normalization of the state and, in addition, it makes explicit the linear relation between $\Phi_j(x|z)$ and $\Phi_{-1-j}(x|z)$ given by
\begin{equation}
\label{reflection}
\Phi_j(x|z) = B(j) \int_{\mathbb C} d^2x' |x-x'|^{4j} \Phi_{-1-j}(x'|z),
\end{equation}
where
\begin{equation}
B(j) = \frac{\nu^{1+2j}}{\pi b^2} \gamma\left(1+b^2(1+2j)\right), \quad\quad \nu = \frac{\pi}{b^2} \gamma\left(1-b^2\right), \quad\quad b^2=(k-2)^{-1},
\end{equation}
with
\begin{equation}
\gamma(x)=\frac{\Gamma(x)}{\Gamma(1-\bar x)}.
\end{equation}

The invariance of correlation functions under the symmetries generated by $J^a_0$, $\bar{J}^a_0$ and $L_n$, $\bar{L}_n$, $n = \pm 1,0$, determines the functional form of the propagator and the three-point function up to certain constants depending strictly on the spin configurations. They were computed in \cite{Teschner:1997ft,Teschner:1999ug}. For the two-point function one has
\begin{equation}
\label{Euclidean2pt}
\left\langle \Phi_{j_1}(x_1|z_1) \Phi_{j_2}(x_2|z_2)\right\rangle = \left[\delta(1+j_1+j_2)\delta^2(x_{12}) + B(j_1)\delta(j_1-j_2) |x_{12}|^{4j_1} \right] |z_{12}|^{-4\Delta_1},
\end{equation}
where $x_{12}=x_2-x_1$ and $z_{12}=z_2-z_1$. The three-point function is expressed as
\begin{equation}
\left\langle \prod_{i=1}^3 \Phi_{j_i}(x_i|z_i) \right\rangle = C(j_a) 
\prod_{\sigma} |x_{\sigma_1\sigma_2}|^{2j_{\sigma}} |z_{\sigma_1\sigma_2}|^{-2\Delta_{\sigma}},
\end{equation}
where the product runs over all cyclic permutations of the labels and we have introduced $j_{\sigma}= j_{\sigma_1}+j_{\sigma_2}-j_{\sigma_3}$ and $\Delta_{\sigma}=\Delta_{\sigma_1}+\Delta_{\sigma_2}-\Delta_{\sigma_3}$. The structure constants $C(j_a) \equiv C(j_1,j_2,j_3)$, as proposed in \cite{Teschner:1997ft}, are given by
\begin{equation}
C(j_a) = \frac{G(1+j_1+j_2+j_3)}{\nu^{-1-j_1-j_2-j_3} G_0}\prod_{\sigma}\frac{G(j_{\sigma})}{G(1+2j_{\sigma_1})},
\end{equation}
where the special function $G(j)$ is constructed by means of the Barnes double gamma function as follows \cite{Zamolodchikov:1995aa},
\begin{equation}
\label{G}
G(j) = b^{-bj\left(b+b^{-1}+bj\right)}\Gamma_2(-bj|b,b^{-1})\Gamma_2(b+b^{-1}+bj|b,b^{-1}),
\end{equation}
and
\begin{equation}
G_0=-2\pi^2\gamma(1+b^2)G(-1).
\end{equation}
The dependence of the three-point function on $(x,\bar x)$ is uniquely fixed by the $SL(2,\mathbb C)$ invariance as long as no $j_{\sigma}$ equals a negative integer for any cyclic permutation $\sigma$. Among the properties of the function $G(j)$ defined by (\ref{G}), the following ones play an important role:
\begin{equation}
G(-1-b^{-2}-j)=G(j),
\end{equation}
\begin{equation}
\label{prop1}
G(-1+j)=\gamma(1+b^2j)G(j),
\end{equation}
\begin{equation}
G(-b^{-2}+j)=b^{2(1+2j)}\gamma(1+j)G(j).
\end{equation}
The relevance of these functional relations rely on the fact that they bring a well-defined meromorphic continuation for $G(j)$ to the whole complex plane, its poles being located at $j=n+mb^{-2}$ and $j=-(n+1)-(m+1)b^{-2}$ for $n,m \in \mathbb N_0$.

The four-point function is expressed as the following integral,
\begin{equation}
\label{4pt}
\left\langle \prod_{i=1}^4 \Phi_{j_i}(x_i|z_i) \right\rangle = \int_{\mathcal P^+} dj C(j_1,j_2,j) B(-1-j) C(j,j_3,j_4) \mathcal G_j(J|X|Z),
\end{equation}
where the integration contour is $\mathcal P^+ = -1/2 + i \lambda$, $\lambda \in \mathbb R_{>0}$, and the nonchiral conformal blocks $\mathcal G_j(J|X|Z)$, $J=(j_1,j_2,j_3,j_4)$, $X=(x_1,x_2,x_3,x_4)$, $Z=(z_1,z_2,z_3,z_4)$, were introduced and exhaustively studied in \cite{Teschner:1999ug}. The decomposition (\ref{4pt}) is valid for 
\begin{equation}
|\mbox{Re}(1+j_1+j_2)|,|\mbox{Re}(j_2-j_1)|<\frac{1}{2},
\end{equation}
and the same for $j_1,j_2 \leftrightarrow j_3,j_4$. Its analytic continuation for other spin configurations is discussed in \cite{Teschner:1999ug,Maldacena:2001km}.

The $SL(2,\mathbb R)$-WZNW model shares with its Euclidean counterpart the affine symmetry though it involves different representations. A key ingredient for generating a consistent spectrum for the model is the so-called spectral flow automorphism which is defined by
\begin{equation}
\label{spectral}
J^3_n \rightarrow \tilde{J}^3_n = J^3_n - \frac{k}{2} \omega \delta_{n,0}, \quad\quad J^{\pm}_n \rightarrow \tilde{J}^{\pm}_n = J^{\pm}_{n \pm \omega},
\end{equation}
with $\omega \in \mathbb Z$, and similarly for the antiholomorphic modes. Notice that the Sugawara-Virasoro algebra is mapped under a spectral flow automorphism into another conformal realization generated by
\begin{equation}
\label{flowVir}
L_n \rightarrow \tilde{L}_n = L_n + \omega J^3_n - \frac{k}{4}\omega^2\delta_{n,0}.
\end{equation}
Unlike in models with underlying compact group symmetries, a spectral flow automorphism generally gives rise to nonequivalent representations when acting on a current module. The spectrum proposed in \cite{Maldacena:2000hw}, for the universal covering group of $SL(2,\mathbb R)$, contains two families of representations of the current algebra: the one extending the principal continuous representations of the universal cover of $SL(2,\mathbb R)$ and their spectral flow images, denoted by $\hat{\mathcal C}^{\alpha,\omega}_{j} \otimes \hat{\mathcal C}^{\alpha,\omega}_{j}$, with $j=-1/2+i\lambda$, $\lambda\in\mathbb R$ and $0 < \alpha \le 1$, and the one extending the lowest-weight discrete series and their spectral flow images, $\hat{\mathcal D}^{+,\omega}_j \otimes \hat{\mathcal D}^{+,\omega}_j$, both types with the same spectral flow number and the same value of $j$ for the left and right sectors. Representations obtained from the highest-weight discrete series can be identified with those built upon the lowest-weight series as
\begin{equation}
\label{seriesid}
\hat{\mathcal D}^{+,\omega}_j=\hat{\mathcal D}^{-,\omega+1}_{-k/2-j},
\end{equation}
allowing us to consider states lying in $\hat{\mathcal D}^{-,\omega}_j \otimes \hat{\mathcal D}^{-,\omega}_j$ as well when computing correlators and restricting the range of values of the spin to the real interval
\begin{equation}
-\frac{k-1}{2}<j<-\frac{1}{2}.
\end{equation}

We shall denote by $\Phi^{\omega}_{jm\bar m}(z)$ the vertex operators associated with the states that are images of the affine primaries under a spectral flow automorphism, where we have introduced the levels $m$ and $\bar m$, with $m-\bar m \in \mathbb Z$ and $m+\bar m \in \mathbb R$, to keep track of the zero modes quantum numbers in the unflowed frame. The OPEs of these operators with the currents can be read from (\ref{spectral}), giving
\begin{equation}
J^{3}(z)\Phi^{\omega}_{jm\bar m}(w) \sim \frac{m+k\omega/2}{z-w}\Phi^{\omega}_{jm\bar m}(w),
\end{equation}
\begin{equation}
 J^{\pm}(z)\Phi^{\omega}_{jm\bar m}(w) \sim \frac{\mp j + m}{(z-w)^{1 \pm \omega}}\Phi^{\omega}_{j,m \pm 1,\bar m}(w),
\end{equation}
and the same for the antiholomorphic currents.

Vertex operators satisfying these OPEs for the case $\omega=0$ can be obtained from the $H_3^+$-WZNW model by performing the following Fourier transformation to the so-called $m$ basis\footnote{Concerning this transformation, let us stress that going from $H^+_3$ to $SL(2,\mathbb{R})$ cannot be reduced to a change of basis, as the Mellin transform can be defined in both cases. It involves an analytic continuation in the space-time coordinates. See \cite{Hosomichi:2001fm} for more details.},
\begin{equation}
\label{fourier}
 \Phi_{jm\bar m}(z) \equiv \Phi^{\omega=0}_{jm\bar m}(z) = \int_{\mathbb C} d^2x x^{j+m} \bar x^{j+ \bar m} \Phi_{-1-j}(x|z).
\end{equation}
Accordingly, correlation functions for these unflowed states are expected to be obtained by using the same transformation on the $H_3^+$-WZNW model correlators, their validity assumed beyond the Euclidean spectrum. Moreover, since correlation functions with the same total spectral flow number just differ in the power dependence on the world-sheet coordinates,  once the conformal dimensions are changed as
\begin{equation}
 \Delta = \Delta_0 -\omega m - \frac{k}{4} \omega^2,
\end{equation}
an expression that follows from (\ref{flowVir}), correlators with unflowed operators as well as those preserving the total flow though involving spectral flowed primaries can be computed by means of this Fourier transform.

The two-point function computed from (\ref{Euclidean2pt}) following this method gives, for $\omega_1+\omega_2=0$,
\begin{eqnarray}
\label{2ptm}
&& \left\langle \Phi^{\omega_1}_{j_1m_1\bar m_1}(z_1) \Phi^{\omega_2}_{j_2m_2\bar m_2}(z_2)\right\rangle = \nonumber \\
&& ~~~~~ ~~~~~ \left[\delta(1+j_1+j_2) + B(-1-j_1) c^{-1-j_1}_{m_1\bar m_1} \delta(j_1-j_2) \right] \delta^{2}(m_1+m_2) z_{12}^{-2\Delta_1}\bar z_{12}^{-2\bar{\Delta}_1},
\end{eqnarray}
where
\begin{equation}
 c^j_{m\bar m}=\frac{\pi}{\gamma(-2j)}\frac{\gamma(-j+m)}{\gamma(1+j+m)},
\end{equation}
and
\begin{equation}
\label{deltacomp}
 \delta^2(m)=\int_{\mathbb C} d^2x x^{m-1} \bar x^{\bar m-1} = 4\pi^2 \delta(m+\bar m)\delta_{m,\bar m}.
\end{equation}
The three-point function with $\omega_1+\omega_2+\omega_3=0$ is given by
\begin{equation}
\label{3ptm}
\left\langle \prod_{i=1}^3 \Phi^{\omega_i}_{j_im_i\bar m_i}(z_i) \right\rangle = C(-1-j_a) W(j_a;m_a,\bar m_a) \delta^2(m_1+m_2+m_3)
\prod_{\sigma} z_{\sigma_1\sigma_2}^{-\Delta_{\sigma}} \bar z_{\sigma_1\sigma_2}^{-\bar{\Delta}_{\sigma}},
\end{equation}
with $W(j_a;m_a,\bar m_a)$ defined as
\begin{equation}
\label{satoh}
W(j_a;m_a,\bar m_a) = \int_{\mathbb C} d^2x_1 d^2x_2 \prod_{\sigma} x_{\sigma_1}^{j_{\sigma_1}+m_{\sigma_1}} \bar x_{\sigma_1}^{j_{\sigma_1}+ \bar m_{\sigma_1}} \left|x_{\sigma_1\sigma_2}\right|^{-2-2j_{\sigma}},
\end{equation}
where $x_3,\bar x_3=1$, and, again, the product runs over all cyclic permutations of the labels. This integral was explicitly computed in terms of certain hypergeometric functions in \cite{Satoh:2001bi}. Concerning spectral flow conserving four-point functions, they were determined from (\ref{4pt}) in \cite{Minces:2007td,Baron:2008qf}. Independent computations of all these correlation functions using the free field approach were performed in \cite{Becker:1993at,Giribet:2000fy,Giribet:2001ft}.

Correlators in the $SL(2,\mathbb R)$-WZNW model can violate the spectral flow number conservation according to the following selection rules,
\begin{eqnarray}
\label{violation}
&& -N_c-N_d+2 \le \sum_{i=1}^{N_c+N_d} w_i \le N_c-2, \quad\quad \mbox{when at least one state is in } \hat{\cal C}^{ w}_{j,\alpha} \otimes \hat{\cal C}^{ w}_{j,\alpha}, \\
&& -N_d+1 \le \sum_{i=1}^{N_c+N_d} w_i \le -1, \quad\quad \mbox{when all states are in } \hat{\cal D}_j^{+,w} \otimes \hat{\cal D}_j^{+,w},
\end{eqnarray}
where $N_c$ is the number of vertex operators associated with states in $\hat{\cal C}^{ w}_{j,\alpha} \otimes \hat{\cal C}^{ w}_{j,\alpha}$ and similarly $N_d$ for those lying in $\hat{\cal D}^{+,w}_j \otimes \hat{\cal D}^{+,w}_j$. By virtue of the series identification (\ref{seriesid}), if states in $\hat{\cal D}_j^{-,w} \otimes \hat{\cal D}_j^{-,w}$ are also taken into account, these rules show that two-, three- and four-point functions can reach maximal violation with $0$, $1$ and $2$ units of spectral flow, respectively. The determination of these spectral flow nonpreserving amplitudes is more involved that those we have already referred. Indeed, a suitable procedure for performing this kind of computation nontrivially includes in the correlators an additional vertex, the so-called spectral flow operator, for each amount of flow violation they present. This method was developed in \cite{Fateev} and it was used in \cite{Maldacena:2001km,Giribet:2001ft} for obtaining the following expression for the three-point function violating the spectral flow conservation in a single unit, namely, for  $\omega_1+\omega_2+\omega_3=\pm 1$:
\begin{equation}
\label{3ptmw}
\left\langle \prod_{i=1}^3 \Phi^{\omega_i}_{j_im_i\bar m_i}(z_i) \right\rangle = C'(-1-j_a)W'(j_a;\pm m_a,\pm \bar m_a) \delta^2(m_1+m_2+m_3\pm k/2)
\prod_{\sigma} z_{\sigma_1\sigma_2}^{-\Delta_{\sigma}} \bar z_{\sigma_1\sigma_2}^{-\bar{\Delta}_{\sigma}},
\end{equation}
where
\begin{equation}
C'(j_a) = \frac{B(j_1) C(-k/2-j_1,j_2,j_3)}{\gamma(-k/2-j_1-j_2-j_3)},
\end{equation}
and
\begin{equation}
\label{Wtilde}
W'(j_a; m_a, \bar m_a) = \sin[\pi(m_3-\bar m_3)] \prod_{i=1}^3 \gamma(1+j_i+ m_i).
\end{equation}
An independent computation of this correlator was performed in \cite{Iguri:2007af} using free field methods. Four-point functions violating the spectral flow conservation are not reported in the literature.

\section{Correlation functions in the space-time picture}
\label{sec3}

The vertex operators in the $SL(2,\mathbb R)$-WZNW model serve as ingredients for the string theory vertex operators describing states created by sources in the boundary of the target space. For example, if $\Phi_j(x|z)$ is a field associated with an unflowed state like in (\ref{fourier}) and $\Theta(z)$ is a spinless world-sheet vertex corresponding to the internal CFT, the sum of their scaling dimensions assumed to equal 1, an operator like
\begin{equation}
V_{j}(x)\sim \int_{\mathbb C} d^2z \Phi_j(x|z) \Theta(z),
\end{equation}
is a vertex describing a string state created by a pointlike source located at $(x,\bar x)$ on the boundary of $AdS_3$, and, by means of the AdS/CFT conjecture, it can be identified with a CFT operator inserted at the same point. Scattering amplitudes involving operators in the space-time representation, unlike amplitudes with states in the $m$ basis like those we have reviewed in the previous section, acquire a similar interpretation when integrated over the moduli space of the string world sheet as correlation functions on the dual two-dimensional CFT.

For unflowed primaries, the definition of the coordinate basis vertex operators comes from the Euclidean model through analytic continuation. The corresponding correlators are the ones of the $H_3^+$-WZNW model. The situation is more complicated when treating spectral flowed primary states since they generally lie in representations with unbounded energy. A solution for this issue was proposed in \cite{Maldacena:2001km}. Let $\left|\Psi\right\rangle$ be an arbitrary lowest-energy state, like any primary in $\hat{\cal C}^{ w}_{j,\alpha}$ or $\hat{\cal D}^{\pm, w}_j$, obeying
\begin{equation}
\tilde{L}_0 \left|\Psi\right\rangle = \Delta_0 \left|\Psi\right\rangle, \quad\quad \tilde{J}^3_0 \left|\Psi\right\rangle = m \left|\Psi\right\rangle, \quad\quad \tilde{J}^{\pm,3}_n \left|\Psi\right\rangle = 0, \quad n=1,2,3,\dots
\end{equation}
The same state can be seen from a spectral flowed frame with $\omega>0$ as satisfying
\begin{equation}
L_0 \left|\Psi\right\rangle = \Delta \left|\Psi\right\rangle, \quad\quad J^3_0 \left|\Psi\right\rangle = \left(m+\frac{k}{2}\omega\right) \left|\Psi\right\rangle, \quad\quad J^{-}_0 \left|\Psi\right\rangle = 0,
\end{equation}
namely, $\left|\Psi\right\rangle$ corresponds to the lowest-weight state of a certain discrete representation $\mathcal D^{+}_J$ of the global $SL(2,\mathbb R)$ algebra generated by the zero modes, the spin being $J=-m-k\omega/2$. Similarly, if the flow number $\omega$ is negative, the spectral flow automorphism turns $\left|\Psi\right\rangle$ into the highest-weight state of a discrete representation $\mathcal D^{-}_J$ with $J=m+k\omega/2$. The $SL(2,\mathbb R)$ algebra generated by $J_0^a$ is identified with the space-time isometries of the background and the global $SL(2)$ symmetries of the CFT at the boundary. Accordingly, vertex operators having flowed primaries and their global descendants as moments were proposed in \cite{Maldacena:2001km} as those relevant for physical applications.

A couple of comments are in place concerning these vertices. On the one hand, the eigenvalues of $\tilde{J}^3_0$ and its antiholomorphic counterpart do not necessarily agree, therefore it is also the case for the global right- and left-moving $SL(2)$ spins, namely, spectral flowed vertices are no longer expected to be spinless operators, their space-time planar spin being given by the difference between $J$ and $\bar J$. This number has to be an integer in order for the corresponding correlation functions to be single-valued. Secondly, since the lowest-weight state of the discrete representation $\mathcal D^{+}_J$ and the highest-weight of $\mathcal D^{-}_J$ both contribute to the same operator, flowed vertices are not labeled by the spectral flow number but its absolute value instead.

We shall denote the flowed vertex operators by $\Phi^{j\omega}_{J\bar J}(x|z)$ where $\omega$ is now the (positive) amount of spectral flow and the superscript $j$ was introduced to remind us the spin of the unflowed states this vertex is built from. The transformation between the $x$ basis and the $m$ basis is carried on as before, namely, we have
\begin{equation}
\label{fourierJM}
 \Phi^{j\omega}_{JM\bar J\bar M}(z) = \int_{\mathbb C} d^2x x^{J+M} \bar x^{\bar J+ \bar M} \Phi^{-1-j,\omega}_{-1-J,-1-\bar J}(x|z),
\end{equation}
where $M$ and $\bar M$ are the eigenvalues of $J^3_0$ and $\bar J^3_0$. Unlike the nontrivial reflection of the global $SL(2)$ spin, the reflection of $j$ in (\ref{fourierJM}) is rather a choice by virtue of (\ref{reflection}). For the extremal-weight moments we have the following identification:
\begin{equation}
\label{identif}
 \Phi^{j\omega}_{J,\pm J,\bar J, \pm\bar J}(z) = \Phi^{\mp\omega}_{jm\bar m}(z), \quad \quad m=\pm J \pm \frac{k}{2}\omega,~~\bar m=\pm \bar J \pm \frac{k}{2}\omega.
\end{equation}
Notice that the integral transform (\ref{fourierJM}) cancels whenever $J-\bar J$ is an integer unless $M-\bar M$ is also an integer number. We will assume this in the following.

Thus far, there are two methods available to perform the computation of correlators involving spectral flowed operators in the space-time picture. According to the first one, the correlation functions are computed directly in the $x$ basis, using 
two alternative integral expressions
 for the flowed vertex, the one manifestly local in space-time and the other local in the world-sheet coordinates, both involving the fusion of an unflowed state and the so-called spectral flow operator $\Phi_{-k/2}(x|z)$. The consistency of these definitions as well as their equivalence were studied in \cite{Maldacena:2001km}. The total spectral flow carried by a correlator translates into the same amount of new insertions in the associated unflowed amplitude. The corresponding Knizhnik-Zamolodchikov equations are very difficult to solve when more than five vertices are considered, despite of the fact that they are simplified by the null descendant conditions derived for the spectral flow operators. The usefulness of the procedure is thus reduced to the determination of two- and certain three-point functions as in \cite{Maldacena:2001km,Minces:2005nb}.

Beyond any computational difficulty, the stringent constraint of this method lies in the fact that the $x$ basis expressions for the flowed operators were originally introduced in \cite{Maldacena:2001km} for states with a single unit of spectral flow, their generalizations for $\omega > 1$ being still lacking. An alternative strategy, the so-called FZZ method introduced in \cite{Fateev}, can be followed in order to bypass this limitation. According to it, one has to begin with an unflowed correlator, insert as many spectral flow operators as units of spectral flow the final correlation function violates, transform to the $m$ basis in order to consistently remove the dependence of the correlator on the ``unphysical'' insertion points and, then, transform back to the space-time basis. The FZZ recipe is a difficult procedure to accomplish since, even if a correlation function is already known in the $m$ basis, its last step is a highly nontrivial task. Nevertheless, in \cite{Maldacena:2001km} it was noticed that Eq.~(\ref{identif}) could serve to finally realize the FZZ strategy in the simplest cases. Since identification (\ref{identif}) holds only for highest- and lowest-weight states, moments with other values for $M$ and $\bar M$ having contributions from flowed affine descendants, an explicit transformation back to the $x$ basis is, in general, unfeasible, forcing the applicability of the procedure to heavily rely on the knowledge of the dependence of the correlators on the boundary coordinates. When computing two- and three-point functions one is able to take advantage of their invariance under the global $SL(2)$ symmetry to determine this dependence up to an overall constant. The method is, therefore, expected to be useful in these cases. Indeed, in \cite{Maldacena:2001km} it was successfully implemented to obtain the propagator for a state in an arbitrary spectral flow sector and the three-point function with two unflowed operators and a single vertex with $\omega=1$. In the next section we shall extend these computations allowing an arbitrary assignment for the amount of spectral flow for each vertex involved. Since some constraints on the spin configurations would, in general, appear it will be convenient to firstly review the simplified case of the two-point function following these lines.

Invariance under the action of the algebra generated by $J^a_0$ and $\bar J^a_0$ implies the following form for the two-point function:
\begin{equation}
\label{2pt}
  \left\langle \Phi_{J\bar J}^{j_1\omega_1}(x_1) \Phi_{J\bar J}^{j_2\omega_2}(x_2) \right\rangle = D(J,\bar J; j_a,\omega_a) x_{12}^{2J} \bar x_{12}^{2\bar J},
\end{equation}
where $D(J,\bar J; j_a,\omega_a)$ is a coefficient not determined by the $SL(2)$ global symmetry. We will omit the explicit dependence on the world-sheet coordinates till the end of the computation.

By means of (\ref{fourierJM}) we can transform this expression to the $m$ basis. Using
\begin{equation}
\label{int}
  \int_{\mathbb C} d^2x |x|^{2a} |1-x|^{2b} x^n (1-x)^m = \pi \frac{\Gamma(1+a+n)\Gamma(1+b+m)\Gamma(-1-a-b)}{\Gamma(-a)\Gamma(-b)\Gamma(2+a+b+n+m)},
	\end{equation}
we obtain,
\begin{equation}
\label{2ptmbasis}
\left\langle \Phi_{JM_1\bar J\bar M_1}^{j_1\omega_1} \Phi_{JM_2\bar J\bar M_2}^{j_2\omega_2} \right\rangle = \delta^2(M_1+M_2) D(-1-J,-1-\bar J; -1-j_a,\omega_a) c^{-1-J}_{-M_1,-\bar M_1},
\end{equation}
and therefore,
\begin{equation}
\label{2ptmbasisext}
   \left\langle \Phi_{J,-J,\bar J,-\bar J}^{j_1,\omega_1} \Phi_{JJ\bar J\bar J}^{j_2,\omega_2} \right\rangle = \pi^2 \frac{D(-1-J,-1-\bar J; -1-j_a,\omega_a)}{|1+2J|^2}.
\end{equation}

By virtue of (\ref{2ptm}) and (\ref{identif}), it follows that $D(J,\bar J; j_a,\omega_a)$ cancels unless $\omega_1=\omega_2$ and
\begin{eqnarray}
  && D(J,\bar J; j_a,\omega_1=\omega_2)=V_{\mbox{\footnotesize conf}}\frac{|1+2J|^2}{\pi^2} \left[\delta(1+j_1+j_2) + \pi \frac{B(j_1)}{\gamma(-2j_1)} \times \right. \nonumber \\
	&& ~~~~~ ~~~~~ \left. \frac{\gamma\left(-1-j_1-J+k\omega_1/2\right)}{\gamma\left(j_1-J+k\omega_1/2\right)} \delta(j_1-j_2) \right],
\end{eqnarray}
where $V_{\mbox{\footnotesize conf}}$ is the conformal volume on the sphere. Introducing
\begin{equation}
\label{lambda}
   \Lambda^{j\omega}_{J\bar J}=c^{j}_{-1-J+k\omega/2,-1-\bar J+k\omega/2} = \frac{\pi}{\gamma(-2j)} \frac{ \gamma_{\omega}\left(-1-j-J\right)}{\gamma_{\omega}\left(j-J\right)},
\end{equation}
with
\begin{equation}
   \gamma_{\omega}(x)=\gamma\left(x+k\omega/2\right),
\end{equation}
we finally get
\begin{eqnarray}
\label{2ptdef}
  && \left\langle \Phi_{J\bar J}^{j_1\omega_1}(x_1|z_1) \Phi_{J\bar J}^{j_2\omega_2}(x_2|z_2) \right\rangle = V_{\mbox{\footnotesize conf}}\frac{|1+2J|^2}{\pi^2} \delta_{\omega_1,\omega_2} \left[\delta(1+j_1+j_2) + \right. \nonumber \\
	&& ~~~~~ ~~~~~ ~~~~~ ~~~~~ ~~~~~ \left. B(j_1) \Lambda^{j_1\omega_1}_{J\bar J} \delta(j_1-j_2) \right] x_{12}^{2J} \bar x_{12}^{2\bar J} z_{12}^{-2\Delta_1} \bar z_{12}^{-2\bar{\Delta}_1},
\end{eqnarray}
where
\begin{equation}
\Delta_1=\Delta_{01}-\omega_1 J - \omega_1 + k\omega_1^2/4.
\end{equation}

Notice that
\begin{equation}
   \lim_{J,\bar J\rightarrow j} \Lambda^{j,\omega=0}_{J\bar J}=\frac{\pi^2}{V_{\mbox{\footnotesize conf}}(1+2j)^2};
\end{equation}
therefore, the regular term of the two-point function for unflowed states is reached in the same limit, namely, by taking 
\begin{equation}
\label{limit}
   \Phi_j(x|z)=\lim_{J,\bar J\rightarrow j} \Phi^{j,\omega=0}_{J\bar J}(x|z).
\end{equation}
The contact term is not expected to be obtained from (\ref{2ptdef}) since the global spins for both insertions agree.

In order to determine the two-point function in space-time, besides of the vertex operators we have considered, the contribution coming from the internal CFT must be taken into account. If we assume that the corresponding vertices are unit normalized and that we are dealing with states obeying the Virasoro constraint, this contribution reduces to a factor $z_{12}^{-2h_1}\bar z_{12}^{-2\bar h_1}$, where $(h_1,\bar h_1)$ are the internal conformal weights. The target space two-point function is obtained after integrating over the moduli space, dividing by the volume of the conformal group on the sphere. We get
\begin{equation}
\label{2pttarget}
\left\langle V_{J\bar J}^{j_1\omega_1}(x_1) V_{J\bar J}^{j_2\omega_2}(x_2) \right\rangle = |1+2J|^2 \delta_{\omega_1,\omega_2} \left[\delta(1+j_1+j_2) + B(j_1) \Lambda^{j_1\omega_1}_{J\bar J} \delta(j_1-j_2) \right] x_{12}^{2J} \bar x_{12}^{2\bar J},
\end{equation}
where
\begin{equation}
J=-1+\frac{k}{4}\omega_1 -\frac{1}{\omega_1}\left(1-h_1-\Delta_{01}\right),
\end{equation}
and a similar expression holds for $\bar J$. Notice that, unlike the situation in \cite{Maldacena:2001km}, we had no need to rescale the vertex operators neither for states built up upon the continuous nor the discrete series in order to obtain the correct propagator. While the two-point function for the former is explicitly finite, it is also the case for the propagator of a short string, since the divergence coming from the evaluation of $\delta(j_1-j_2)$ at $j_1=j_2$ in (\ref{2pttarget}) is canceled by a pole developed by $\Gamma\left(j_1-J+k\omega_1/2\right)$ in the denominator of (\ref{lambda}). Even though, there is a subtlety when performing this cancellation, since it gives rise to an additional factor $|1+2j-(k-2)\omega|$. In turn this factor brings the $|1+2j|$ needed in order to properly reproduce the factorization of the four-point function onto the short string with $\omega=0$.
%
%
\subsection{Notation} 
In the unflowed frame, our notation differs from the one used in \cite{Maldacena:2001km} simply in the signs of the spins, namely,
\begin{equation}
\label{notation1}
\Phi_{j}(x|z) \leftrightarrow \tilde{\Phi}_{j}(x|z)=\Phi_{-j}(x|z).
\end{equation}
A little more care should be taken when $\omega\ne 0$. Since the definition of the Fourier transform connecting the $x$- and the $m$ basis in \cite{Maldacena:2001km} is different from (\ref{fourierJM}), for recovering the expressions obtained in \cite{Maldacena:2001km} from ours, the following identification must be imposed,
\begin{equation}
\label{notation}
\Phi^{j\omega}_{J\bar J}(x|z) \leftrightarrow \tilde{\Phi}^{j\omega}_{J\bar J}(x|z)=\frac{1}{V_{\mbox{\footnotesize conf}}} \int_{\mathbb C} d^2x' (x-x')^{-2J} (\bar x-\bar x')^{-2\bar J} \Phi^{-j,\omega}_{-1+J,-1+\bar J}(x'|z),
\end{equation}
since the integration is needed in order to adjust the dependence of the correlation functions in the space-time coordinates. Indeed, using (\ref{int}) twice, it is straightforward to get
\begin{eqnarray}
\label{2ptmaldnot}
  && \left\langle \tilde{\Phi}_{J\bar J}^{j_1\omega_1}(x_1|z_1) \tilde{\Phi}_{J\bar J}^{j_2\omega_2}(x_2|z_2) \right\rangle = \frac{1}{V_{\mbox{\footnotesize conf}}} \delta_{\omega_1,\omega_2} \left[\delta(1-j_1-j_2) + \right. \nonumber \\
	&& ~~~~~ ~~~~~ ~~~~~ ~~~~~ ~~~~~ \left. B(-j_1) \Lambda^{-j_1,\omega_1}_{-1+J,-1+\bar J} \delta(j_1-j_2) \right] x_{12}^{-2J} \bar x_{12}^{2\bar J} z_{12}^{-2\Delta_1} \bar z_{12}^{-2\bar{\Delta}_1},
\end{eqnarray}
where, now, $\Delta_1=\Delta_{01}-\omega_1 J + k\omega_1^2/4$, in full agreement with the corresponding correlator obtained in \cite{Maldacena:2001km}.

Notice that, apart from the change in the signs of the spins and an eventual overall factor, the correspondence (\ref{notation}) resembles the reflection identity (\ref{reflection}), a symmetry that it is not clear to be present in the spectral flowed frame. In the next section we shall make extensive use of (\ref{notation}) when checking the agreement of our results with other correlators reported in the literature.

\section{Three-point correlators}
\label{sec4}

Some three-point functions have been already computed in the space-time picture. A correlator involving one vertex with a single unit of spectral flow and two unflowed operators was obtained in \cite{Maldacena:2001km} using the two strategies we have already described in the previous section, and the three-point function involving two operators in the $\omega=1$ sector and an unflowed vertex and certain three-point functions with all of their insertions carrying a unit of flow were determined in \cite{Minces:2005nb} directly in the $x$ basis. Other three-point functions were also computed in \cite{Cardona:2009hk} within the context of the superstring theory on $AdS_3 \times S^3 \times T^4$. In this section we extend these computations for an arbitrary amount of spectral flow.

The invariance under the global symmetry generated by the zero modes of the $SL(2)$ algebra establishes the following space-time dependence for the three-point function,
\begin{equation}
\label{3ptgeneral}
\left\langle \Phi_{J_1\bar J_1}^{j_1\omega_1}(x_1) \Phi_{J_2\bar J_2}^{j_2\omega_2}(x_2) \Phi_{J_3\bar J_3}^{j_3\omega_3}(x_3) \right\rangle = D(J_a,\bar J_a;j_a,\omega_a) \prod_{\sigma} x_{\sigma_1\sigma_2}^{J_{\sigma}} \bar x_{\sigma_1\sigma_2}^{\bar J_{\sigma}},
\end{equation}
where the product runs over all cyclic permutations of the labels, $J_{\sigma}=J_{\sigma_1}+J_{\sigma_2}-J_{\sigma_3}$, $D(J_a,\bar J_a;j_a,\omega_a)$ is a constant to be determined and, again, we are omitting any dependence on the world-sheet coordinates until the final expressions.

After transforming to the $m$ basis according to (\ref{fourierJM}), we obtain
\begin{eqnarray}
\label{3ptJM}
&& \left\langle \Phi_{J_1M_1\bar J_1\bar M_1}^{j_1\omega_1} \Phi_{J_2M_2\bar J_2\bar M_2}^{j_2\omega_2} \Phi_{J_3M_3\bar J_3\bar M_3}^{j_3\omega_3} \right\rangle = \nonumber \\
&& ~~~~~ ~~~~~ D(-1-J_a,-1-\bar J_a;-1-j_a,\omega_a) W(J_a,\bar J_a;M_a,\bar M_a) \delta^2(M_1+M_2+M_3),
\end{eqnarray}
where $W(J_a,\bar J_a;M_a,\bar M_a)$ is the generalization of (\ref{satoh}) for $J_a \ne \bar J_a$, namely,
\begin{equation}
\label{satoh2}
W(J_a,\bar J_a;M_a,\bar M_a) = \int_{\mathbb C} d^2x_1 d^2x_2 \prod_{\sigma} x_{\sigma_1}^{J_{\sigma_1}+M_{\sigma_1}} \bar x_{\sigma_1}^{\bar J_{\sigma_1}+ \bar M_{\sigma_1}} x_{\sigma_1\sigma_2}^{-1-J_{\sigma}} \bar x_{\sigma_1\sigma_2}^{-1-\bar J_{\sigma}},
\end{equation}
with $x_3,\bar x_3=1$. Unlike (\ref{satoh}), there is no explicit formula for this integral, but for our purposes it will be enough to compute it when one of the insertions, say the first one, corresponds to a lowest-weight state, namely, for $M_1=-J_1$ and $\bar M_1=-\bar J_1$. In this case, Eq.~(\ref{satoh2}) simplifies to
\begin{equation}
\label{satoh2b}
W_1(J_a,\bar J_a;M_2,\bar M_2) = \int_{\mathbb C} d^2x_1 d^2x_2 x_{2}^{J_{2}+M_{2}} \bar x_{2}^{\bar J_{2}+ \bar M_{2}} \prod_{\sigma} x_{\sigma_1\sigma_2}^{-1-J_{\sigma}} \bar x_{\sigma_1\sigma_2}^{-1-\bar J_{\sigma}},
\end{equation}
and it can be explicitly solved using (\ref{int}) twice. We obtain
\begin{equation}
\label{satoh2bb}
W_1(J_a,\bar J_a;M_2,\bar M_2) =  \frac{\pi^2 \gamma(-1-J_1-J_2-J_3)\gamma(1+J_2+M_2)\gamma(-J_{31})\gamma(-J_{12})}{\gamma(-2 J_1)\gamma(-J_{31}-J_2+M_2)},
\end{equation}
where we are now following the standard notation, writing $J_{\sigma_1\sigma_2}$ instead of $J_{\sigma}$.

\subsection{Three-point function with a single flowed insertion.}

The case of the three-point function with only one spectral flowed operator was previously treated in \cite{Maldacena:2001km}. We recall it here for completeness.

Let us assume that the flowed vertex is inserted in the first point. According to (\ref{violation}), we necessarily should have $\omega_1=1$, so that (\ref{3ptJM}) reduces to
\begin{eqnarray}
\label{wequal1}
&& \left\langle \Phi_{J_1,-J_1,\bar J_1,-\bar J_1}^{j_1,\omega_1=1} \Phi_{j_2m_2\bar m_2} \Phi_{j_3m_3\bar m_3} \right\rangle = \nonumber \\
&& ~~~~~ ~~~~~ D(-1-J_a,-1-\bar J_a;-1-j_a,\omega_a) W_1(J_a,\bar J_a;m_2,\bar m_2) \delta^2(-J_1+m_2+m_3),
\end{eqnarray}
where $J_{i}=\bar J_{i}=j_{i}$, $i=2,3$. Setting $m_1=-J_1-k/2$ and $\bar m_1=-\bar J_1-k/2$, it follows from (\ref{3ptmw}), (\ref{Wtilde}) and (\ref{identif}), that
\begin{eqnarray}
&& D(-1-J_a,-1-\bar J_a;-1-j_a,\omega_a) = \frac{C'(-1-j_a)W'(j_a;m_a,\bar m_a)}{W_1(J_a,\bar J_a;m_2,\bar m_2)} = \nonumber \\
&& ~~~~~ ~~~~~ \frac{C'(-1-j_a) \gamma(1+j_1-J_1-k/2) \gamma(-2J_1)}{\pi^2 \gamma(-1-J_1-j_2-j_3)\gamma(-J_{31})\gamma(-J_{12})}.
\end{eqnarray}
Since $\Phi_{j_2m_2\bar m_2}(x_2|z_2)$ and $\Phi_{j_3m_3\bar m_3}(x_3|z_3)$ run over all the allowed representations, $m_2$, $\bar m_2$, $m_3$ and $\bar m_3$ can take arbitrary values. The delta functions appearing in (\ref{3ptmw}) and (\ref{wequal1}) do not imply any constraint to the spin configurations and, therefore, they have been canceled. 
Therefore,
\begin{eqnarray}
\label{3ptx1}
&& \left\langle \Phi_{J_1\bar J_1}^{j_1,\omega_1=1}(x_1|z_1) \Phi_{j_2}(x_2|z_2) \Phi_{j_3}(x_3|z_3) \right\rangle = \nonumber \\
&& \frac{C'(j_a) \gamma(1-j_1+J_1-k/2) \gamma(2+2J_1)}{\pi^2 \gamma(2+J_1+j_2+j_3)\gamma(1+J_{31})\gamma(1+J_{12})}\prod_{\sigma} x_{\sigma_1\sigma_2}^{J_{\sigma}} \bar x_{\sigma_1\sigma_2}^{\bar J_{\sigma}} z_{\sigma_1\sigma_2}^{-\Delta_{\sigma}} \bar z_{\sigma_1\sigma_2}^{-\bar{\Delta}_{\sigma}}.
\end{eqnarray}

The target space correlator is obtained after integrating over the moduli space, once those contributions to the scaling dimensions coming from the internal space are taken into account. One gets, up to a proper normalization,
\begin{eqnarray}
\label{3ptx1target}
&& \left\langle V_{J_1\bar J_1}^{j_1,\omega_1=1}(x_1) V_{j_2}(x_2) V_{j_3}(x_3) \right\rangle = \nonumber \\
&& \frac{C'(j_a) \gamma(1-j_1+J_1-k/2) \gamma(2+2J_1)}{\pi^2 \gamma(2+J_1+j_2+j_3)\gamma(1+J_{31})\gamma(1+J_{12})}\prod_{\sigma} x_{\sigma_1\sigma_2}^{J_{\sigma}} \bar x_{\sigma_1\sigma_2}^{\bar J_{\sigma}}.
\end{eqnarray}

Using the following identity,
\begin{eqnarray}
\label{integralmalda}
&& \frac{\gamma(c)}{\pi \gamma(b)\gamma(c-b)} \left|x^{1-c}\right|^2 \int_{\mathbb C} d^2y \left| y^{b-1} (x-y)^{c-b-1} (1-y)^{-a} \right|^2 = \nonumber \\
&& ~~~~~ ~~~~~ \left| {}_2F_1\left.\left(
\begin{array}{c}
	a,b \\
	c
\end{array} \right| x \right) \right|^2 - \lambda \left| x^{1-c} {}_2F_1\left.\left(
\begin{array}{c}
	1+b-c,1+a-c \\
	2-c
\end{array} \right| x \right) \right|^2,
\end{eqnarray}
where ${}_2F_1$ represents the hypergeometric function and
\begin{equation}
\lambda=\frac{\gamma(c)^2\gamma(a-c+1)\gamma(b-c+1)}{(c-1)^2\gamma(a)\gamma(b)},
\end{equation}
together with the fact that
\begin{equation}
\label{hyper}
{}_2F_1\left.\left(
\begin{array}{c}
	a,b \\
	a
\end{array} \right| x \right)=(1-x)^{-b},
\end{equation}
it is possible to show from (\ref{3ptx1}), that 
\begin{eqnarray}
\label{3ptx1notvie}
&& \left\langle \tilde{\Phi}_{J_1\bar J_1}^{j_1,\omega_1=1}(x_1|z_1) \tilde{\Phi}_{j_2}(x_2|z_2) \tilde{\Phi}_{j_3}(x_3|z_3) \right\rangle = \nonumber \\
&& \frac{ B(j_1) C(-k/2+j_1,-j_2,-j_3)}{V_{\mbox{\footnotesize conf}}} \frac{\gamma(j_1+J_1-k/2)\gamma(J_{23})}{\gamma(j_1+j_2+j_3-k/2)} \prod_{\sigma} x_{\sigma_1\sigma_2}^{-J_{\sigma}} \bar x_{\sigma_1\sigma_2}^{-\bar J_{\sigma}} z_{\sigma_1\sigma_2}^{-\Delta_{\sigma}} \bar z_{\sigma_1\sigma_2}^{-\bar{\Delta}_{\sigma}},
\end{eqnarray}
with $\Delta_1=\Delta_{01}- J_1 + k/4$. This expression coincides with the corresponding three-point function reported in \cite{Maldacena:2001km}.

\subsection{Three-point functions with two flowed insertions}

When two spectral flowed vertex operators are inserted in a three-point function, the expression (\ref{satoh2b}) can be further simplified since another extremal-weight state must be considered in order to use the identity (\ref{identif}). In our case we shall consider a highest-weight state in the second point so that $M_2=J_2$ and $\bar M_2=\bar J_2$. Introducing
\begin{equation}
\label{satohast}
W_2(J_a,\bar J_a) \equiv W_1(J_a,\bar J_a;J_2,\bar J_2) =\frac{\pi^2 \gamma(-1-J_1-J_2-J_3)\gamma(1+2J_2)\gamma(-J_{12})}{\gamma(-2 J_1)},
\end{equation}
Eq.~(\ref{3ptJM}) reduces to
\begin{eqnarray}
\label{wequal2}
&& \left\langle \Phi_{J_1,-J_1,\bar J_1,-\bar J_1}^{j_1\omega_1} \Phi_{J_2J_2\bar J_2\bar J_2}^{j_2\omega_2} \Phi_{j_3m_3\bar m_3} \right\rangle = \nonumber \\
&& ~~~~~ ~~~~~ D(-1-J_a,-1-\bar J_a;-1-j_a,\omega_a) W_2(J_a,\bar J_a) \delta^2(-J_1+J_2+m_3),
\end{eqnarray}
where $J_3=\bar J_3=j_3$.

According to the selection rules for the spectral flow violation (\ref{violation}), there are just two cases of interest: the spectral flow conserving correlator and the correlation function violating the spectral flow conservation in a unit. Both cases must be treated separately.

\subsubsection{Spectral flow preserving correlator}

Let us assume that the first two insertions correspond to the flowed vertex operators and that $\omega_1=\omega_2$. By virtue of (\ref{identif}) we can identify $\Phi_{J_1,-J_1,\bar J_1,-\bar J_1}^{j_1\omega_1}$ in (\ref{wequal2}) with $\Phi^{\omega_1}_{j_1m_1\bar m_1}$, where $m_1=-J_1-k\omega_1/2$ and $\bar m_1=-\bar J_1-k\omega_1/2$, and $\Phi_{J_2J_2\bar J_2\bar J_2}^{j_2\,\omega_2}$ with $\Phi^{-\omega_2}_{j_2m_2\bar m_2}$, where $m_2=J_2+k\omega_2/2$ and $\bar m_2=\bar J_2+k\omega_2/2$. Under these identifications, we can equate (\ref{3ptm}) with (\ref{wequal2}), obtaining
\begin{eqnarray}
\label{wequal2b}
&& D(-1-J_a,-1-\bar J_a;-1-j_a,\omega_a)  = \frac{C(-1-j_a) W(j_a;m_a,\bar m_a)}{W_2(J_a,\bar J_a)} = \nonumber \\
&& ~~~~~ ~~~~~ \frac{C(-1-j_a) W(j_a;m_a,\bar m_a) \gamma(-2 J_1)}{\pi^2 \gamma(-1-J_1-J_2-j_3)\gamma(1+2J_2)\gamma(-J_{12})}.
\end{eqnarray}
For writing (\ref{wequal2b}) we have canceled the delta functions appearing in (\ref{3ptm}) and (\ref{wequal2}) since $m_3$ and $\bar m_3$ take arbitrary values and, as before, these delta functions do not imply any constraint to the spin assignments.

It follows that
\begin{equation}
\label{3ptconsx}
\left\langle \Phi_{J_1\bar J_1}^{j_1\omega_1}(x_1|z_1) \Phi_{J_2\bar J_2}^{j_2\omega_2}(x_2|z_2) \Phi_{j_3}(x_3|z_3) \right\rangle = C(j_a)\widehat W(j_a;J_a,\bar J_a) \prod_{\sigma} x_{\sigma_1\sigma_2}^{J_{\sigma}} \bar x_{\sigma_1\sigma_2}^{\bar J_{\sigma}} z_{\sigma_1\sigma_2}^{-\Delta_{\sigma}} \bar z_{\sigma_1\sigma_2}^{-\bar{\Delta}_{\sigma}},
\end{equation}
where we have introduced
\begin{equation}
\label{widehat}
\widehat W(j_a;J_a,\bar J_a) = \frac{W(-1-j_a;m_a,\bar m_a) \gamma(2+2 J_1)}{\pi^2 \gamma(2+J_1+J_2+j_3)\gamma(-1-2J_2)\gamma(1+J_{12})},
\end{equation}
with $m_1=1+ J_1- k\omega_1/2$, $\bar m_1=1+\bar J_1-k\omega_1/2$, $m_2=-1- J_2+ k\omega_2/2$ and $\bar m_2=-1-\bar J_2+k\omega_2/2$.

The target space correlation function is computed, as usual, by integrating over the moduli space, after adjusting the world-sheet dependence of the fields by considering their scaling dimensions in the internal space $(h_i, \bar h_i)$. We get
\begin{equation}
\label{3ptconsxtarget}
\left\langle V_{J_1\bar J_1}^{j_1\omega_1}(x_1) V_{J_2\bar J_2}^{j_2\omega_2}(x_2) V_{j_3}(x_3) \right\rangle = C(j_a)\widehat W(j_a;J_a,\bar J_a) \prod_{\sigma} x_{\sigma_1\sigma_2}^{J_{\sigma}} \bar x_{\sigma_1\sigma_2}^{\bar J_{\sigma}},
\end{equation}
with
\begin{equation}
J_{i}=-1+\frac{k}{4}\omega_{i} -\frac{1}{\omega_{i}}\left(1-h_{i}-\Delta_{0i}\right), \qquad i=1,2.
\end{equation}

There are several interesting consistency checks that can be performed on (\ref{3ptconsx}). The simplest one corresponds to the limit (\ref{limit}). Setting $\omega_i=0$ and taking the limit $J_{i},\bar J_{i}\rightarrow j_{i}$ for $i=1,2$, we get $W(-1-j_a;m_a,\bar m_a) \rightarrow W_2(-1-j_a,-1-j_a)$. By virtue of (\ref{satohast}), all the gamma functions appearing in (\ref{widehat}) are canceled and $\widehat W(j_a;J_a,\bar J_a) \rightarrow 1$. At the end, the three-point function for unflowed vertices is reproduced. 

Another check follows through (\ref{notation}) when $\omega_1=\omega_2=1$. Indeed, using (\ref{integralmalda}) twice together with (\ref{hyper}) it is straightforward to prove that (\ref{3ptconsx}) reduces to
\begin{eqnarray}
\label{3ptconsxmald}
&& \left\langle \tilde{\Phi}_{J_1\bar J_1}^{j_1\omega_1}(x_1|z_1) \tilde{\Phi}_{J_2\bar J_2}^{j_2\omega_2}(x_2|z_2) \tilde{\Phi}_{j_3}(x_3|z_3) \right\rangle = \nonumber \\
&& ~~~~~ ~~~~~ \frac{C(-j_a) W(-j_a;m_a,\bar m_a)}{V_{\mbox{\footnotesize conf}}^2} \prod_{\sigma} x_{\sigma_1\sigma_2}^{-J_{\sigma}} \bar x_{\sigma_1\sigma_2}^{-\bar J_{\sigma}} z_{\sigma_1\sigma_2}^{-\Delta_{\sigma}} \bar z_{\sigma_1\sigma_2}^{-\bar{\Delta}_{\sigma}},
\end{eqnarray}
where $\Delta_{i}=\Delta_{0i}-\omega_{i}J_{i}+k\omega^2_{i}/4$ with $i=1,2$, $m_1=-J_1-kw_1/2$, $\bar m_1=-\bar J_1-kw_1/2$, $m_2=J_2+kw_2/2$, $\bar m_2=\bar J_2+kw_2/2$, . Setting $\omega_1=\omega_2=1$ one recovers the corresponding expression found in \cite{Minces:2005nb}. 

More interestingly, inserting an identity at the third point in (\ref{3ptconsx}), the emergence of both terms of the two-point function (\ref{2ptdef}) can be proven. Due to the factor $G^{-1}(1+2\epsilon)$ coming from $C(j_a)$ when setting $j_3 \equiv \epsilon \rightarrow 0$, the correlator (\ref{3ptconsx}) vanishes unless further terms behave singular in the same limit. When $j_1 \sim j_2$, there are two such factors: $G(j_1-j_2+\epsilon)$ and $G(j_2-j_1+\epsilon)$. The singular behavior of the first one can be represented, by virtue of (\ref{prop1}), as $-b^{-2}(j_1-j_2+\epsilon)^{-1} G(-1)$. Similarly, we have $G(j_2-j_1+\epsilon)\sim -b^{-2}(j_2-j_1+\epsilon)^{-1} G(-1)$. It follows that
\begin{equation}
\lim_{\epsilon\rightarrow 0} \frac{G(j_1-j_2+\epsilon)G(j_2-j_1+\epsilon)}{G(-1)G(1+2\epsilon)} = 2\pi\gamma(b^2) \delta(j_1-j_2),
\end{equation}
since
\begin{equation}
\label{deltaaprox}
\lim_{\epsilon\rightarrow 0} \frac{2\epsilon}{x^2-\epsilon^2} = - 2\pi \delta(x),
\end{equation}
and, therefore,
\begin{equation}
\label{first}
C(j_1,j_2,\epsilon) \longrightarrow B(j_1)\delta(j_1-j_2).
\end{equation}
On the other hand, we have
\begin{eqnarray}
\label{second}
&& W(-1-j_a;m_a,\bar m_a) \sim \nonumber \\
&& ~~~~~ \int_{\mathbb C} d^2x_1 d^2x_2 |x_1|^{-2j_1+2J_1-k\omega_1} \bar x_1^{\bar J_1-J_1} |x_2|^{-4-2j_1-2J_2+k\omega_1} \bar x_2^{J_2-\bar J_2} |x_{12}|^{4j_1-2\epsilon} \sim \nonumber \\
&& ~~~~~ \delta^2(J_1-J_2) \frac{\pi \gamma_{\omega_1}(-1-j_1-J_1)}{\gamma(-2j_1+\epsilon)\gamma_{\omega_1}(j_1-J_1-\epsilon)} \longrightarrow \delta^2(J_1-J_2) \Lambda^{j_1 \omega_1}_{J_1\bar J_1},
\end{eqnarray}
where we have used (\ref{deltacomp}) and (\ref{int}). We recover the second term of (\ref{2ptdef}) after inserting (\ref{first}) and (\ref{second}) into (\ref{3ptconsx}). The overall factor $V_{\mbox{\footnotesize conf}}$ follows from the evaluation of the delta function.

In order to obtain the first term of (\ref{2ptdef}) one has to have a little more care. When $1+j_1+j_2\sim 0$, we find just one singular factor in $C(j_a)$, namely, $G(1+j_1+j_2+\epsilon)$, which behaves as $-b^{-2}(1+j_1+j_2+\epsilon)^{-1} G(-1)$, so that
\begin{equation}
\label{justone}
C(j_a) \sim -\frac{G(j_1-j_2)G(j_2-j_1)}{2\pi^2 v^{-1-j_1-j_2}G(1+2j_1)G(1+2j_2)} \frac{2\epsilon}{1+j_1+j_2+\epsilon}.
\end{equation}
Unlike the case where only unflowed vertex operators are considered, there is no singular behavior coming from the dependence of (\ref{3ptconsx}) in the space-time coordinates but, again, it comes from the integral $W(-1-j_a;m_a,\bar m_a)$. Indeed, we have
\begin{eqnarray}
\label{justtwo}
&& W(-1-j_a;m_a,\bar m_a) = \int_{\mathbb C} d^2x_1 d^2x_2 |x_1|^{-2j_1+2J_1-k\omega} \bar x_1^{\bar J_1-J_1} |x_2|^{-4-2j_2-2J_2+k\omega} \times \nonumber \\
&& ~~~ \bar x_2^{J_2-\bar J_2} |x_{12}|^{2j_1+2j_2-2\epsilon} |1-x_1|^{2j_2-2j_1} |1-x_2|^{2j_1-2j_2} \sim \frac{\pi\delta^2(J_1-J_2-1-j_1-j_2)}{1+j_1+j_2-\epsilon},
\end{eqnarray}
where we have used $|x_{12}|^{2(j_1+j_2-\epsilon)} \sim \pi (1+j_1+j_2-\epsilon)^{-1} \delta^2(x_{12})$. By virtue of (\ref{deltaaprox}), we get
\begin{equation}
\label{CW}
C(j_a) W(-1-j_a;m_a,\bar m_a) \longrightarrow \delta(1+j_1+j_2) \delta^2(J_1-J_2).
\end{equation}
We recover the first term of (\ref{2ptdef}) after replacing this expression in (\ref{3ptconsx}). Again, the factor $V_{\mbox{\footnotesize conf}}$ arises from the evaluation of the delta function in (\ref{CW}).

\subsubsection{Spectral flow violating correlator}

Let us assume, again, that the first two insertions are those with flowed vertex operators. Without any loss of generality, we can set $\omega_1=\omega_2+1$. If we now identify $\Phi_{J_1,-J_1,\bar J_1,-\bar J_1}^{j_1\omega_1}$ in (\ref{wequal2}) with $\Phi^{\omega_2+1}_{j_1m_1\bar m_1}$, where $m_1=-J_1-k\omega_2/2-k/2$ and $\bar m_1=-\bar J_1-k\omega_2/2-k/2$, and $\Phi_{J_2J_2\bar J_2\bar J_2}^{j_2\omega_2}$ with $\Phi^{-\omega_2}_{j_2m_2\bar m_2}$, where $m_2=J_2+k\omega_2/2$ and $\bar m_2=\bar J_2+k\omega_2/2$, we can equate (\ref{3ptmw}) with (\ref{wequal2}). We obtain,
\begin{equation}
\label{wequal2c}
D(-1-J_a,-1-\bar J_a;-1-j_a,\omega_a)  = \frac{C'(-1-j_a) W'(j_a;m_a,\bar m_a) \gamma(-2 J_1)}{\pi^2 \gamma(-1-J_1-J_2-J_3)\gamma(1+2J_2)\gamma(-J_{12})}.
\end{equation}
Notice that, again, we have been able to cancel the delta functions appearing in (\ref{3ptmw}) and (\ref{wequal2}) since $m_3$ and $\bar m_3$ can acquire any value.

Replacing (\ref{wequal2c}) in (\ref{3ptgeneral}) we get the following expression for the spectral flow violating three-point function,
\begin{eqnarray}
\label{3ptconsxb}
&& \left\langle \Phi_{J_1\bar J_1}^{j_1\omega_1}(x_1|z_1) \Phi_{J_2\bar J_2}^{j_2\omega_2}(x_2|z_2) \Phi_{j_3}(x_3|z_3) \right\rangle = \nonumber \\
&& ~~~~~ ~~~~~ C'(j_a)\widehat W'(j_a;J_a,\bar J_a)   \prod_{\sigma} x_{\sigma_1\sigma_2}^{J_{\sigma}} \bar x_{\sigma_1\sigma_2}^{\bar J_{\sigma}} z_{\sigma_1\sigma_2}^{-\Delta_{\sigma}} \bar z_{\sigma_1\sigma_2}^{-\bar{\Delta}_{\sigma}}
\end{eqnarray}
%
where we have introduced
\begin{equation}
\label{hatW'}
\widehat W'(j_a;J_a,\bar J_a) = \frac{\gamma(2+2 J_1) \gamma(2+2 J_2) \gamma_{\omega_2}(-1-j_2-J_2)}{\pi^2 \gamma(2+J_1+J_2+j_3)\gamma(1+J_{31})\gamma(1+J_{12})\gamma_{\omega_1}(j_1-J_1)}
\end{equation}

Notice that, after setting $\omega_2=0$, in the limit $J_2,\bar J_2 \rightarrow j_2$ we obtain 
\begin{equation}
\widehat W'(j_a;J_a,\bar J_a) \rightarrow  \frac{\gamma(1-j_1+J_1-k/2) \gamma(2+2J_1)}{\pi^2 \gamma(2+J_1+j_2+j_3)\gamma(1+J_{31})\gamma(1+J_{12})}
\end{equation}
so that we recover (\ref{3ptx1}), as expected.

The target space correlator is given by
\begin{eqnarray}
\label{3ptconxdeftarget}
&& \left\langle V_{J_1\bar J_1}^{j_1\omega_1}(x_1) V_{J_2\bar J_2}^{j_2\omega_2}(x_2) V_{j_3}(x_3) \right\rangle = C'(j_a)\widehat W'(j_a;J_a,\bar J_a) \prod_{\sigma} x_{\sigma_1\sigma_2}^{J_{\sigma}} \bar x_{\sigma_1\sigma_2}^{\bar J_{\sigma}}.
\end{eqnarray}

Let us finally quote the expression of the three-point function following the notation of \cite{Maldacena:2001km}. By virtue of (\ref{notation}), using (\ref{integralmalda}) twice and (\ref{hyper}), we get
\begin{eqnarray}
\label{3ptconxmalda}
&& \left\langle \tilde{\Phi}_{J_1\bar J_1}^{j_1\omega_1}(x_1|z_1) \tilde{\Phi}_{J_2\bar J_2}^{j_2\omega_2}(x_2|z_2) \tilde{\Phi}_{j_3}(x_3|z_3) \right\rangle = \nonumber \\
&& \frac{ C'(-j_a) \gamma_{\omega_2}(j_2-J_2)\gamma(J_{23})}{V_{\mbox{\footnotesize conf}}^2\gamma_{\omega_1}(1-j_1-J_1)} \prod_{\sigma} x_{\sigma_1\sigma_2}^{-J_{\sigma}} \bar x_{\sigma_1\sigma_2}^{-\bar J_{\sigma}} z_{\sigma_1\sigma_2}^{-\Delta_{\sigma}} \bar z_{\sigma_1\sigma_2}^{-\bar{\Delta}_{\sigma}},
\end{eqnarray}
where $\Delta_{i}=\Delta_{0i}-\omega_{i}J_{i}+k\omega^2_{i}/4$, $i=1,2$.

\subsection{Three-point functions with three flowed insertions}

When there are three flowed insertions in a correlator, another highest-weight state should be also considered in the third point, so that we can set $M_3=J_3$ and $\bar M_3=\bar J_3$. Unlike in the previous cases, the FZZ recipe fails to give the most general correlation function since the delta functions in (\ref{3ptm}), (\ref{3ptmw}) and (\ref{3ptJM}) now do impose some constraints on the possible values of the spins. Indeed, the method is restraint to be valid only when the arguments of these delta functions vanish.

Under these nontrivial conditions,  Eq. (\ref{3ptJM}) reduces to
\begin{eqnarray}
\label{wequal3}
&& \left\langle \Phi_{J_1,-J_1,\bar J_1,-\bar J_1}^{j_1\omega_1} \Phi_{J_2J_2\bar J_2\bar J_2}^{j_2\omega_2} \Phi_{J_3J_3\bar J_3\bar J_3}^{j_3\omega_3} \right\rangle = \nonumber \\
&& ~~~~~ ~~~~~ D(-1-J_a,-1-\bar J_a;-1-j_a,\omega_a) W_3(J_a,\bar J_a) \delta^2(-J_1+J_2+J_3),
\end{eqnarray}
where $W_3(J_a,\bar J_a)$ is the restriction of $W_2(J_a,\bar J_a)$ on the hyperplanes $J_3=J_1-J_2$, $\bar J_3=\bar J_1-\bar J_2$, {{\em i.e.},
\begin{equation}
\label{satohast2}
W_3(J_a,\bar J_a) = \frac{\pi^2}{|1+2J_1|^2}.
\end{equation}

As before, the spectral flow preserving correlator and the correlation function violating the spectral flow conservation in a unit must be treated separately.

\subsubsection{Spectral flow conserving correlator}

Without any loss of generality, we can sort the states inside the correlation function in descending order of spectral flow, namely, we can freely assume that $\omega_1 \ge \omega_2\ge \omega_3$. For a spectral flow preserving correlator we accordingly have $\omega_1 = \omega_2+ \omega_3$. By virtue of (\ref{identif}) we can identify $\Phi_{J_1,-J_1,\bar J_1,-\bar J_1}^{j_1\omega_1}$ in (\ref{wequal3}) with $\Phi^{\omega_1}_{j_1m_1\bar m_1}$, where $m_1=-J_1-k\omega_1/2$ and $\bar m_1=-\bar J_1-k\omega_1/2$, $\Phi_{J_2J_2\bar J_2\bar J_2}^{j_2\omega_2}$ with $\Phi^{-\omega_2}_{j_2m_2\bar m_2}$, where $m_2=J_2+k\omega_2/2$ and $\bar m_2=\bar J_2+k\omega_2/2$, and $\Phi_{J_3J_3\bar J_3\bar J_3}^{j_3\omega_3}$ with $\Phi^{-\omega_3}_{j_3m_3\bar m_3}$, where $m_3=J_3+k\omega_3/2$ and $\bar m_3=\bar J_3+k\omega_3/2$, and under these identifications we can equate (\ref{3ptm}) with (\ref{wequal3}), obtaining
\begin{eqnarray}
\label{wequal3b}
&& D(-1-J_a,-1-\bar J_a;-1-j_a,\omega_a)  = \frac{C(-1-j_a) W(j_a;m_a,\bar m_a)}{W_3(J_a,\bar J_a)} 
\end{eqnarray}
The delta functions appearing in (\ref{3ptm}) and (\ref{wequal3}) have been canceled, but it must be kept in mind that (\ref{wequal3b}) is only valid when their arguments vanish.

It follows that
\begin{eqnarray}
\label{3ptconsx,3w}
&&\left\langle \Phi_{J_1\bar J_1}^{j_1\omega_1}(x_1|z_1) \Phi_{J_2\bar J_2}^{j_2\omega_2}(x_2|z_2) \Phi_{J_3\bar J_3}^{j_3\omega_3}(x_3|z_3) \right\rangle =  \nonumber \\
&& ~~~~~ ~~~~~ \frac{|1+2J_1|^2}{\pi^2} \, C(j_a) W(-1-j_a;m_a,\bar m_a) \prod_{\sigma} x_{\sigma_1\sigma_2}^{J_{\sigma}} \bar x_{\sigma_1\sigma_2}^{\bar J_{\sigma}} z_{\sigma_1\sigma_2}^{-\Delta_{\sigma}} \bar z_{\sigma_1\sigma_2}^{-\bar{\Delta}_{\sigma}},
\end{eqnarray}
with $m_1=1+ J_1- k\omega_1/2$, $\bar m_1=1+\bar J_1-k\omega_1/2$, $m_2=-1- J_2+ k\omega_2/2$ and $\bar m_2=-1-\bar J_2+k\omega_2/2$. The restrictions on the spins are
\begin{equation}
\label{restrict}
J_1=J_2+J_3+1 \quad \mbox{and} \quad \bar J_1=\bar J_2+\bar J_3+1.
\end{equation}
The target space correlation function is obtained, as before, by omitting the world-sheet dependence.

\subsubsection{Spectral flow violating correlator}

If, again, we sort the states inside a correlator in descending order of spectral flow, the condition for maximally violating the spectral flow number conservation is achieved by imposing $|\omega_1-\omega_2-\omega_3|=1$. By virtue of (\ref{identif}) we can make the same identifications for the states as in the previous case and equate (\ref{3ptmw}) with (\ref{wequal3}). We finally obtain
\begin{eqnarray}
\label{3ptconsx,3wb}
&&\left\langle \Phi_{J_1\bar J_1}^{j_1\omega_1}(x_1|z_1) \Phi_{J_2\bar J_2}^{j_2\omega_2}(x_2|z_2) \Phi_{J_3\bar J_3}^{j_3\omega_3}(x_3|z_3) \right\rangle =  \nonumber \\
&& ~~~~~ ~~~~~ \frac{|1+2J_1|^2}{\pi^2} \, C'(j_a) \prod_{i=1}^3 \gamma(-j_i+ m_i) \prod_{\sigma} x_{\sigma_1\sigma_2}^{J_{\sigma}} \bar x_{\sigma_1\sigma_2}^{\bar J_{\sigma}} z_{\sigma_1\sigma_2}^{-\Delta_{\sigma}} \bar z_{\sigma_1\sigma_2}^{-\bar{\Delta}_{\sigma}},
\end{eqnarray}
with $m_1=1+ J_1- k\omega_1/2$, $\bar m_1=1+\bar J_1-k\omega_1/2$, and $m_{i}=-1- J_{i}+ k\omega_{i}/2$, $\bar m_{i}=-1-\bar J_{i}+k\omega_{i}/2$, $i=2,3$.
As before, this expression is valid as long as Eq.~(\ref{restrict}) holds. The target space correlation function is reached after eliminating the world-sheet dependence.

\section{Concluding remarks}

In this paper we have computed three-point functions on the sphere for the $SL(2,\mathbb R)$-WZNW model following a procedure based on the FZZ recipe \cite{Fateev}, similar to the one discussed in \cite{Maldacena:2001km}. We have considered spectral flow preserving and nonpreserving correlation functions involving vertex operators defined on arbitrary spectral flow frames. The Fou  rier-like transformation for going from the space-time picture to the $m$ basis slightly differs from the one given in \cite{Maldacena:2001km}. It allowed us to avoid singularities and to unify the treatment of short and long string states. Several consistency checks were performed on the expressions obtained. They reduce to the known correlators in the unflowed limit, and they all agree with correlation functions previously referred to in the literature when setting $\omega=1$.

For those cases in which at least one insertion corresponds to an unflowed state, no restriction is found for the spins involved. By contrast, when all three vertices belong to a nontrivial spectral flow sector, the FZZ recipe fails to give the most general correlator and only constrained spin configurations are allowed. It would be interesting to further explore the model in order to improve the method for solving this issue.

\section*{Acknowledgments}

We are grateful to G.~Giribet, C.~N\'u\~nez and A.~Solotar for carefully reading the manuscript. We also thank M.~Schvellinger, R.~Ferraro and E.~Herscovich for stimulating discussions. This work has been partially supported by the Projects No. UBACyT 20020100100669 and No. PIP 11220080100507.

\end{document}